\documentclass[11pt]{article}
\usepackage[letterpaper,margin=0.9in]{geometry}
\usepackage{amsmath,amssymb,mathtools}
\usepackage{graphicx}
\usepackage{booktabs}
\usepackage{array}
\usepackage{enumitem}
\usepackage{algorithm}
\usepackage{algpseudocode}
\usepackage{fancyhdr}
\usepackage{url}
\usepackage[hidelinks]{hyperref}
\usepackage{xcolor}
\usepackage{listings}
\usepackage{microtype}

\setlist[itemize]{leftmargin=1.6em,itemsep=0.2em,topsep=0.2em}
\setlist[enumerate]{leftmargin=1.8em,itemsep=0.25em,topsep=0.25em}
\newcommand{\cl}{\operatorname{cl}}
\newcommand{\LPM}{\operatorname{LPM}}
\newcommand{\HW}{\operatorname{HW}}
\newcommand{\Cov}{\operatorname{Cov}}
\newcommand{\CertifiedHits}{\operatorname{CertifiedHits}}
\newcommand{\bankPenalty}{\operatorname{bankPenalty}}

\begin{document}
\thispagestyle{plain}

\begin{center}
{\LARGE\bfseries Certified Bank-Aware Fast-Path Coverage for\\[0.2em]
Hardware-Constrained NDN Forwarding Lookup\par}
\vspace{0.35em}
{\large Extended Preprint: Algorithms, Combinatorial Structure, and Design-Space Characterization\par}
\vspace{2.2em}
{\large\bfseries Amirreza Khorasanian\par}
{\large Electrical Engineering Student, ECE Department, University of Tehran\par}
{\large \texttt{akh5793@gmail.com}\par}
\vspace{1.5em}
{\large August 2026\par}
\end{center}

\vspace{1em}
\begin{center}\textbf{Abstract}\end{center}

Named Data Networking (NDN) replaces address-based forwarding with longest-prefix lookup over variable-length hierarchical names. This creates a difficult fit with bounded hardware fast paths: parser depth is finite, trie state consumes scarce memory, memory can be banked, and a shallow match is correct only when it preserves the action of the true longest-prefix match (LPM).

We formulate the problem as \emph{certified fast-path coverage}. An offline mapper chooses a prefix-closed set of parser-visible trie nodes and a separate set of action-equivalence certificates under a shared object-equivalent budget. A certificate relates a selected shallow FIB prefix to a deeper true-LPM class carrying the same forwarding action; applying the shortcut requires a sound class guard, and all uncertain traffic falls back to an exact path. We present BPC-Greedy, a feasibility-filtered, certificate-aware placement heuristic, and BPC-Portfolio, a multi-strategy wrapper spanning exact-heavy, repair-based, and shallow-shortcut constructions.

This extended preprint develops the algorithmic and combinatorial structure that was compressed in the conference-format manuscript. In particular, it shows that (i) prefix placement is a tree-closure knapsack problem with reusable ancestor cost, (ii) for a fixed selected prefix set and uniform certificate cost, certificate selection is exactly solved by a top-$k$ rule because candidate certificate classes partition eligible trace traffic, (iii) bank assignment is a separate load-partition feasibility layer, and (iv) the difficult coupling arises because changing the selected prefix set changes both the deepest hardware match and the certificate opportunity set. We also describe the exact memory-only dynamic program, the bank-repair local search, and every branch of the portfolio.

On a synthetic parser-depth sweep and a 50k replay of public NDN testbed names under a reproducible synthetic forwarding-action snapshot, modeled certified coverage increases from 46.48\% for the best exact-prefix-only bank-aware baseline to 64.06\% at total budget $M=48$. These numbers characterize placement opportunity under the stated model; the current evaluator uses the known true LPM as an oracle-style guard, certificate-table bank traffic is not modeled, and no claim is made about synthesized latency, area, or throughput. The extended analysis makes these boundaries explicit and identifies a natural next problem: jointly synthesizing cheap one-sided guards for high-value action-equivalent classes.

\textbf{Preprint note.} This manuscript extends the conference-format version with additional algorithmic analysis, worked examples, implementation correspondence, and explicit discussion of architectural boundaries. The reported numerical results are unchanged.

\newpage
\section{Introduction}
Named Data Networking (NDN) replaces host-address-based forwarding with name-based forwarding \cite{ndnarch,ndnpkt}. Instead of routing toward fixed-width destination addresses, an NDN forwarder processes Interests for hierarchical names such as \texttt{/edu/ucla/video/lecture/seg5}. This naming model enables content-centric retrieval, in-network caching, and Interest aggregation, but it also changes the forwarding lookup problem: the router must parse a variable number of components and compute longest-prefix matches over a hierarchical namespace.

Hardware pipelines, by contrast, are intentionally bounded. They expose finite parser depth, fixed stage counts, finite memory capacity, restricted memory bandwidth, and constrained update paths. A fast-path design must therefore answer a question that is partly algorithmic and partly architectural: which forwarding decisions are worth representing in hardware, and which of those decisions can be accepted without changing the exact forwarding action?

A naive answer is to store popular exact FIB prefixes. This is safe, but it can spend one object per exact-LPM class and additionally pay structural trie-closure cost. A more aggressive answer is to store a shallower prefix that covers a large subtree. That is cheap, but potentially wrong: the subtree may contain deeper prefixes whose forwarding actions override the shallow action. The central tension is therefore not merely compression; it is semantics-preserving compression.

We study this tension through action-equivalence certificates. If a selected shallow prefix $s$ and a deeper true-LPM class $t$ have the same forwarding action, then using $s$ is semantically safe for traffic whose exact class is $t$. The certificate records this relation. A separate sound guard is required to prove that the current packet belongs to $t$; packets that cannot be certified fall back. This makes the accepted fast path conservative by construction.

The main algorithm, BPC-Greedy, does not first optimize exact-prefix placement and then bolt certificates on afterward. Instead, when it considers adding a prefix closure, it evaluates the certificate-aware marginal gain of that closure. A shallow prefix can therefore be selected even when its own exact-LPM traffic is modest, provided it enables high-weight same-action descendant classes. BPC-Portfolio then treats several structurally different heuristics as complementary search biases and returns the best feasible candidate.

The extended preprint focuses on the combinatorial structure underlying this mapper. Its main additions are:
\begin{itemize}
\item a precise decomposition into tree-closure placement, certificate selection, bank assignment, and budget splitting;
\item an exact dynamic program for the memory-only prefix-closed subproblem and an explanation of the bank-repair local search;
\item a proof that top-$k$ certificate selection is optimal for fixed $S$ under the current unit-cost trace model;
\item a detailed account of why the full objective is nevertheless nonseparable: adding a prefix changes the deepest selected match and therefore changes which certificate pairs exist;
\item a visual and algorithmic explanation of the BPC-Portfolio branches;
\item a more explicit account of the oracle guard, the bank abstraction, and the path toward guard-aware physical cost models.
\end{itemize}

\begin{figure}[t]
\centering
\includegraphics[width=0.86\linewidth]{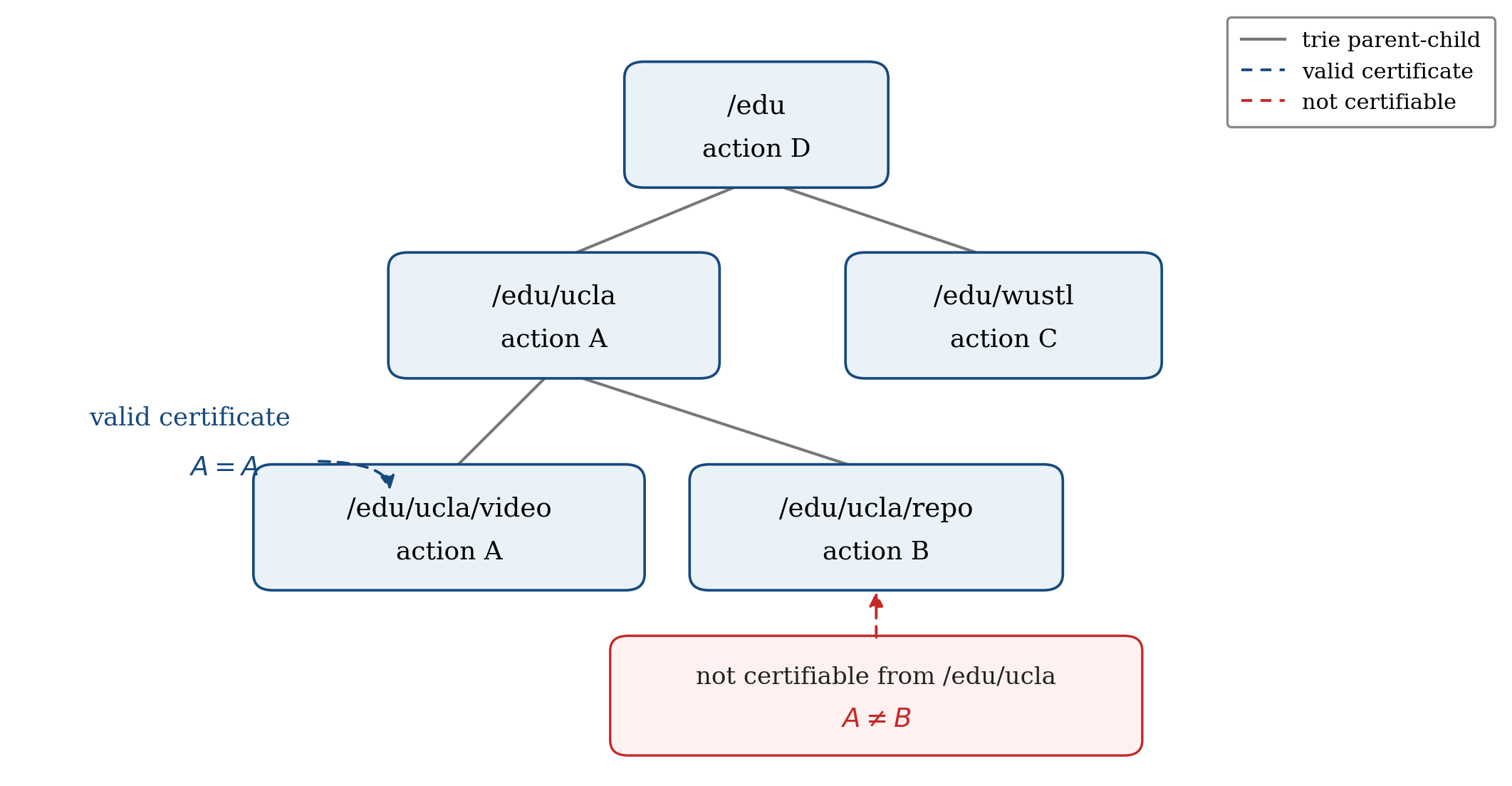}
\caption{Running example. A shallow selected prefix may safely stand in for some deeper exact-LPM classes but not others. Certificates encode same-action relations; they do not relax correctness.}
\label{fig:running}
\end{figure}

\section{NDN Lookup and a Running Example}
An NDN forwarder maintains multiple data structures, including a Content Store (CS), Pending Interest Table (PIT), and Forwarding Information Base (FIB) \cite{lisurvey,nfd}. This paper isolates the FIB lookup subproblem. Each FIB prefix $p$ has an action $a(p)$, representing an outgoing face or next-hop set.

Consider the FIB in Fig.~\ref{fig:running}. The shallow prefix \texttt{/edu/ucla} and deeper prefix \texttt{/edu/ucla/video} both map to action A, whereas \texttt{/edu/ucla/repo} maps to B. Therefore the same shallow state is safe for one descendant class and unsafe for another.

A trace provides workload weight. For example, suppose 65 Interests have true LPM \texttt{/edu/ucla/video}, 20 have true LPM \texttt{/edu/ucla/repo}, and 15 have true LPM \texttt{/edu/wustl}. The FIB determines what forwarding semantics exist; the trace determines how often each semantic class is exercised. The mapper uses both.

\section{Model and Notation}
Let $T$ be a weighted trace of Interest names. Each name $x=(c_1,\ldots,c_{\ell(x)})$ has weight $w(x)$. Let $P$ be the FIB prefixes embedded in a prefix trie with node set $V$. We write $p\preceq x$ when prefix $p$ matches $x$ component-wise.

The exact forwarding class is
\begin{equation}
\LPM(x)=\arg\max_{p\in P,\;p\preceq x}|p|.\tag{1}
\end{equation}
with exact action $a^\star(x)=a(\LPM(x))$.

The mapper selects a parser-visible, prefix-closed set $S\subseteq V$. Structural closure nodes may belong to $S$ without being FIB entries. Only $S\cap P$ carries actions. With parser depth $D$, the deepest selected match is
\begin{equation}
\HW(x,S)=\arg\max_{s\in S,\;s\preceq x,\;|s|\le D}|s|.\tag{2}
\end{equation}
If no selected node matches, $\HW(x,S)=\bot$.

Let $K$ be a set of certificates. A certificate $(s,t)$ requires
\begin{equation}
s\in S\cap P,\qquad t\in P,\qquad s\preceq t,\qquad a(s)=a(t).\tag{3}
\end{equation}
The target $t$ need not be selected and certificates themselves have no prefix-closure requirement. This is important: the purpose of a certificate is precisely to avoid storing every deeper target as selected trie state.

The object-equivalent budget is
\begin{equation}
|S|+\gamma|K|\le M,\tag{4}
\end{equation}
where $\gamma$ is the normalized certificate cost. The main experiment uses $\gamma=1$ as a reference normalization, not as a measured equality of silicon area.

Selected trie state is assigned to $B$ abstract banks. If $b(s)$ is the bank of selected node $s$ and $\lambda(s)$ is its modeled load, then
\begin{equation}
\sum_{s\in S:b(s)=j}\lambda(s)\le C,\qquad j=1,\ldots,B.\tag{5}
\end{equation}
In the current implementation, $\lambda(s)$ is the trace-derived true-LPM count for action-bearing FIB nodes and zero for structural closure nodes. Certificates are not charged bank load. Section~\ref{sec:banks} discusses why this is a feasibility proxy rather than a runtime memory model.

A target-class guard $g_t(x)$ must satisfy the one-sided soundness condition
\begin{equation}
g_t(x)=1\;\Rightarrow\;\LPM(x)=t.\tag{6}
\end{equation}
False negatives are allowed: a guard may decline to certify a packet and send it to fallback. False positives are not allowed because they could change the forwarding action.

A packet is certified when either the selected match is exact or a stored same-action certificate is accompanied by a sound target-class guard:
\begin{equation}
\operatorname{certified}(x,S,K)=
\begin{cases}
1, & \HW(x,S)=\LPM(x),\\
1, & \exists t:(\HW(x,S),t)\in K\land g_t(x)=1,\\
0, & \text{otherwise.}
\end{cases}\tag{7}
\end{equation}
The primary metric is
\begin{equation}
\Cov(T,S,K)=\frac{\sum_{x\in T}w(x)\operatorname{certified}(x,S,K)}{\sum_{x\in T}w(x)}.\tag{8}
\end{equation}

\section{Combinatorial Anatomy of the Placement Problem}
\begin{figure}[t]
\centering
\includegraphics[width=0.78\linewidth]{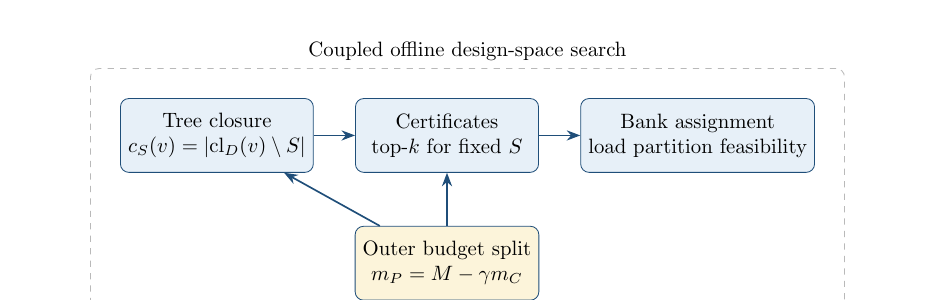}
\caption{Combinatorial decomposition. Certificate selection is easy once $S$ is fixed; the difficulty is that changing $S$ changes closure cost, deepest selected matches, certificate opportunities, and bank feasibility.}
\label{fig:decomp}
\end{figure}

The full mapper is easier to understand when decomposed into four coupled layers: tree closure, certificate choice, bank partitioning, and budget splitting. Fig.~\ref{fig:decomp} summarizes the decomposition.

\subsection{Prefix closure creates reusable path cost}
Let $\cl_D(v)$ denote the parser-visible ancestor closure of candidate FIB prefix $v$. The incremental object cost of adding $v$ to a current selection $S$ is
\begin{equation}
c_S(v)=|\cl_D(v)\setminus S|.\tag{9}
\end{equation}

This is not a fixed item cost. If some ancestors are already selected, the same deep prefix becomes cheaper.

\textbf{Lemma 1 (Ancestor reuse).} If $S\subseteq S'$, then $c_{S'}(v)\le c_S(v)$ for every candidate $v$.

\emph{Proof.} Because $S\subseteq S'$, removing $S'$ from the same closure can only remove at least as many already-present nodes as removing $S$. \hfill$\square$

This tiny observation has a large algorithmic consequence: a static ranking such as ``traffic divided by depth'' cannot capture the true marginal cost. Previously selected ancestors create reuse, and the greedy score must be recomputed after every commit.

\subsection{Memory-only exact placement is a tree-knapsack dynamic program}
The repository contains an exact dynamic program, PCT-DP, for the memory-only prefix-closed subproblem. For a selected node $v$, let $F_v(k)$ be the maximum exact-LPM weight obtainable from the subtree of $v$ with $k$ selected nodes, conditioned on selecting $v$ (the root is free). If $w_v$ is $v$'s FIB weight or zero for a structural node, then the recurrence is a tree convolution:
\begin{equation}
F_v(k)=w_v+\max_{k_1+\cdots+k_r=k-1}\sum_{i=1}^{r}\max\{0,F_{u_i}(k_i)\}.\tag{10}
\end{equation}
where $u_1,\ldots,u_r$ are children of $v$ and the zero option means not selecting a child subtree. The implementation realizes this recurrence by repeatedly merging child budget tables.

PCT-DP is useful conceptually because it separates two facts. Prefix closure alone is structured enough for exact dynamic programming on a tree; adding bank assignment and certificate-aware state-dependent value destroys that clean one-dimensional budget structure.

\subsection{For fixed $S$, certificate selection is exactly top-$k$}
Fix selected state $S$ and parser depth $D$. Every trace packet has a unique deepest selected match $s=\HW(x,S)$ and a unique true LPM $t=\LPM(x)$. If $s\ne t$ and $a(s)=a(t)$, the packet contributes to exactly one candidate pair $(s,t)$.

Define the weight of a pair as
\begin{equation}
q_S(s,t)=\sum_{x\in T:\;\HW(x,S)=s,\;\LPM(x)=t}w(x),\tag{11}
\end{equation}
for action-valid pairs only.

\textbf{Proposition 1 (Optimal certificate selection for fixed $S$).} Under uniform certificate cost and a budget of at most $m_C$ certificates, selecting the $m_C$ pairs with largest $q_S(s,t)$ maximizes certificate hits for the fixed prefix set $S$.

\emph{Proof.} Each eligible trace packet contributes to exactly one pair because both $\HW(x,S)$ and $\LPM(x)$ are unique. Therefore the traffic covered by distinct certificate pairs is disjoint, and total certificate hits are the sum of selected pair weights. Under equal cost, the maximum-weight subset of cardinality $m_C$ is exactly the top-$m_C$ weights. \hfill$\square$

This is one of the useful structural simplifications in the implementation. The certificate subproblem does not require another approximate solver once $S$ is known.

\subsection{The hard coupling: $S$ changes the certificate universe}
Although fixed-$S$ certificate choice is simple, $q_S$ itself changes when $S$ changes. If a deeper selected prefix is added, the deepest selected match for some packets moves from an ancestor $s$ to the newly selected prefix $s'$. Existing pairs $(s,t)$ can disappear, new pairs $(s',t)$ can appear, and some traffic can become exact. Thus certificate value is not an independent item value attached permanently to a FIB node.

This is why BPC-Greedy evaluates a candidate by replaying its joint exact-plus-certificate effect. The relevant marginal gain is
\begin{equation}
\Delta(v\mid S)=\CertifiedHits(T,S\cup\cl_D(v),K^\star_{S\cup\cl_D(v)})-\CertifiedHits(T,S,K^\star_S),\tag{12}
\end{equation}
where $K^\star_S$ is the optimal top-$m_C$ certificate set for fixed $S$.

\subsection{Bank assignment is a separate load-partition problem}
For a candidate selected set, the implementation sorts selected nodes by decreasing $\lambda(s)$ and places each next item into the currently least-loaded bank, then checks Eq.~(5). This is an LPT-style load-balancing heuristic. BankRepairOnly adds a local repair layer that can remove low-density selected leaves when memory or bank constraints are violated.

The bank layer therefore acts as a feasibility filter on a selected prefix set; it is not part of the certificate top-$k$ proof and it is not a simulation of the packet lookup schedule.

\subsection{The generalized optimum contains exact-only placement as a special case}
\textbf{Proposition 2 (Model-level dominance of the generalized feasible region).} Let $\mathrm{OPT}_{\mathrm{exact}}$ be the best feasible coverage with $K=\emptyset$, and let $\mathrm{OPT}_{\mathrm{cert}}$ be the optimum of the generalized model with certificates allowed under the same resource semantics. Then $\mathrm{OPT}_{\mathrm{cert}}\ge\mathrm{OPT}_{\mathrm{exact}}$.

\emph{Proof.} Every exact-only feasible solution remains feasible in the generalized problem by choosing the same $S$, $b$ and $K=\emptyset$. \hfill$\square$

This is a statement about the optimization model, not the heuristic. BPC-Greedy has no approximation guarantee and can in principle miss a better exact-only construction; BPC-Portfolio partly mitigates that risk by explicitly including exact-heavy candidates.

\section{Algorithms}
\subsection{Exact and bank-aware baselines}
\textbf{TopK.} A popularity-only sanity baseline ranks FIB prefixes by trace weight without reasoning about bank pressure or certificate opportunities. It is intentionally simple and is not the headliner baseline.

\textbf{PCT-DP.} The exact memory-only dynamic program described above maximizes exact-prefix weight subject to parser depth, budget, and prefix closure. It serves as a useful oracle for small/memory-only settings.

\textbf{BankGreedy.} BankGreedy scores a candidate using its exact FIB-weight gain divided by missing closure cost plus a bank-balance penalty. Candidates that cannot be feasibly assigned to banks are skipped.

\textbf{BankRepairOnly.} The main exact-only hardware baseline begins from BankGreedy and performs conservative local search. For each candidate closure, if the trial exceeds the node budget or bank capacity, the algorithm repeatedly removes low-density selected leaves that are not part of the new closure. Restricting removal to leaves preserves prefix closure. Density is
\begin{equation}
\rho(v)=\frac{\operatorname{FIBWeight}(v)}{\max\{\lambda(v),1\}}.\tag{13}
\end{equation}
The repaired trial is accepted only when it is feasible and improves exact-prefix value.

\subsection{BPC-Greedy}
BPC-Greedy is the main proposed heuristic. The released code and raw logs retain the historical label BPC-Repair; the final name BPC-Greedy matches the current behavior because a bank-infeasible BPC trial is skipped rather than repaired.

The outer loop tries certificate budget options $m_C\in\Omega$. For each split,
\begin{equation}
m_P=M-\gamma m_C.\tag{14}
\end{equation}
The inner greedy loop starts from $S=\emptyset$ and evaluates every parser-visible FIB prefix. For candidate $v$, it computes missing closure $\Delta_v=\cl_D(v)\setminus S$, rejects a trial exceeding $m_P$, rejects a bank-infeasible trial, exactly chooses the top-$m_C$ certificate pairs for the trial, and measures the resulting increase in certified hits.

\begin{algorithm}[t]
\caption{BankRepairOnly: exact-prefix local repair}
\begin{algorithmic}[1]
\State $S\gets \textsc{BankGreedy}(M,D,B,C)$
\Repeat
\State $S_{best}\gets S$
\ForAll{parser-visible FIB candidates $v$}
\State $R\gets S\cup\cl_D(v)$
\While{$|R|>M$}
\State remove lowest-density selected leaf not in $\cl_D(v)$
\EndWhile
\While{$R$ is bank-infeasible}
\State remove lowest-density selected leaf not in $\cl_D(v)$
\EndWhile
\If{$R$ is feasible and exact value improves}
\State $S_{best}\gets R$
\EndIf
\EndFor
\State $S\gets S_{best}$
\Until{no improvement}
\State \Return $S$
\end{algorithmic}
\end{algorithm}

The implemented score is
\begin{equation}
\operatorname{score}(v\mid S)=\frac{\Delta(v\mid S)}{|\Delta_v|+0.15\,\bankPenalty+\varepsilon}.\tag{15}
\end{equation}
The coefficient 0.15 is heuristic; no approximation ratio is claimed. The best positive-gain trial is committed, the certificate set is recomputed, and the process repeats until no feasible positive-gain candidate remains. Finally, the best budget split is retained.

\subsection{Why BPC is not ``exact greedy + certificates''}
This distinction is central. An exact-first heuristic scores a prefix by traffic for which that prefix is itself the true LPM. BPC scores a prefix by the total safe traffic enabled after certificates are recomputed. A shallow prefix can therefore win even if it has little direct exact traffic.

For example, consider a selected candidate $s$ with three descendant true-LPM classes $t_1,t_2,t_3$ of weights 40, 25, and 20, with actions A, A, B while $a(s)=A$. With two certificate slots, $s$ can enable 65 certificate hits through $(s,t_1)$ and $(s,t_2)$, while $t_3$ remains fallback. Exact-only scoring would not see this value unless it stores $t_1$ and $t_2$ themselves.

\subsection{BPC-Portfolio as multi-bias search}
\begin{figure}[t]
\centering
\includegraphics[width=0.82\linewidth]{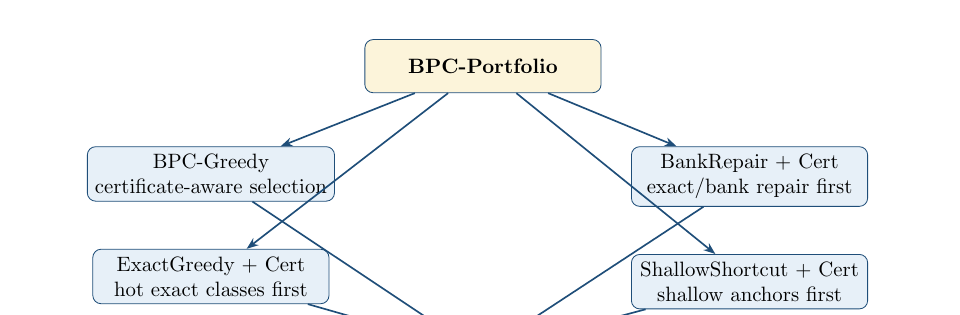}
\caption{BPC-Portfolio is a robust multi-strategy wrapper, not a separate forwarding mechanism. Each branch represents a different search bias; all candidates are validated under the same budget, closure, parser, certificate, and bank rules.}
\label{fig:portfolio}
\end{figure}

BPC-Portfolio evaluates structurally different candidate generators under the same final validation. Fig.~\ref{fig:portfolio} shows the branches.

\begin{algorithm}[t]
\caption{BPC-Greedy}
\begin{algorithmic}[1]
\Require trie $G$, trace $T$, total budget $M$, parser depth $D$, banks $B$, capacity $C$, certificate cost $\gamma$, split set $\Omega$
\State $best\gets$ empty feasible solution
\ForAll{$m_C\in\Omega$}
\State $m_P\gets M-\gamma m_C$; $S\gets\emptyset$
\State evaluate current certified hits with top-$m_C$ certificates
\Repeat
\State $bestCandidate\gets\emptyset$
\ForAll{parser-visible FIB prefixes $v$}
\State $\Delta_v\gets\cl_D(v)\setminus S$
\State $S'\gets S\cup\Delta_v$
\If{$|S'|>m_P$ or $S'$ is bank-infeasible}
\State \textbf{continue}
\EndIf
\State $K'\gets$ top-$m_C$ action-valid pairs induced by $S'$
\State $gain\gets\CertifiedHits(T,S',K')-\CertifiedHits(T,S,K)$
\If{$gain>0$}
\State score trial using Eq.~(15)
\EndIf
\State retain highest-scoring trial
\EndFor
\If{no positive-gain trial exists}
\State \textbf{break}
\EndIf
\State commit best trial and recompute top-$m_C$ certificates
\Until{no improvement}
\State update best if this split has higher certified coverage
\EndFor
\State \Return best
\end{algorithmic}
\end{algorithm}

\textbf{BPC-Greedy branch.} Jointly values prefix closures by certificate-aware marginal safe coverage. This is the branch selected by the portfolio in every reported real-name configuration.

\textbf{BankRepair+Cert.} Reserves $m_C$ certificate slots, uses BankRepairOnly to construct a strong exact-prefix solution under the reduced prefix budget $m_P$, then adds the best certificates available to that fixed $S$. Certificates do not influence which prefixes are selected.

\textbf{ExactGreedy+Cert.} Sorts FIB prefixes by exact trace weight, greedily inserts closures while they fit the prefix budget, then attaches certificates. This branch intentionally favors concentrated exact-LPM traffic.

\textbf{ShallowShortcut+Cert.} Sorts candidates primarily by shallow depth and secondarily by descendant FIB-weight mass. It deliberately seeks cheap ancestors that may act as certificate anchors for large same-action regions. It is the most shortcut-biased portfolio branch.

The portfolio uses lexicographic tie-breaking: higher certified coverage, then higher exact prefix value, then fewer selected nodes.

\subsection{Complexity and why the current implementation is expensive}
Let $n=|P_D|$ be the parser-visible FIB prefix count, $N=|T|$ the trace size, $h$ maximum relevant name depth, $r=|\Omega|$ the number of budget splits, and $I$ the number of committed greedy rounds. A straightforward candidate evaluation scans the trace to compute certificate-aware coverage and performs a bank assignment for the trial. Ignoring small constants, a conservative implementation-level bound is
\begin{equation}
O\!\left(rIn\left(Nh+|S|\log|S|\right)\right),\tag{16}
\end{equation}
which explains why the full 50k BPC sweep is an offline design-space exploration rather than a runtime algorithm.

Several speedups are natural: cache ancestor closures, precompute exact LPM labels, maintain certificate-pair counts incrementally, use lazy-greedy upper bounds, prune dominated budget splits, and compress action-homogeneous regions. These are algorithm-engineering opportunities rather than changes to the semantic model.

\section{The Guard Abstraction and the Next Architectural Problem}
The guard is the most important unresolved architectural component. The current trace evaluator knows $t=\LPM(x)$ and therefore instantiates $g_t(x)$ oracle-style. This is useful for measuring placement opportunity, but it does not establish that a low-cost runtime guard exists.

The key distinction is that a practical guard need not solve full multi-class LPM. It may be a conservative one-sided classifier. Full LPM asks which of many possible prefixes is deepest. A guard for target class $t$ asks only whether there is sufficient evidence to safely certify $t$; uncertainty is allowed to fall back.

\begin{figure}[t]
\centering
\includegraphics[width=0.72\linewidth]{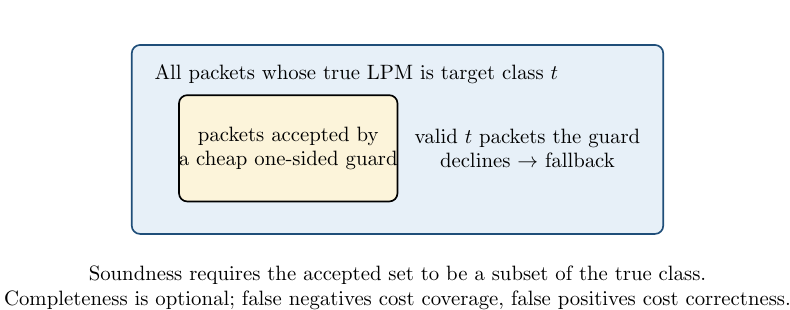}
\caption{One-sided guard interpretation. A useful future guard can conservatively certify an easy subset of a profitable exact-LPM class without classifying every packet.}
\label{fig:guard}
\end{figure}

\begin{table}[t]
\centering
\caption{Implementation correspondence.}
\small
\begin{tabular}{p{0.19\linewidth}p{0.31\linewidth}p{0.42\linewidth}}
\toprule
Concept & File/function & Role\\
\midrule
Trie and exact semantics & \path{ndnhw/trie.py} & path insertion, exact \texttt{lpm()}, deepest selected match, ancestor closure\\
Memory-only exact DP & \path{ndnhw/pct_dp.py::pct_dp} & exact prefix-closed dynamic program without banks/certificates\\
Bank greedy & \path{ndnhw/bank_greedy.py} & exact-value greedy with closure and bank feasibility\\
Exact local repair & \path{ndnhw/bank_repair.py} & add closure, remove low-density leaves until memory/bank feasible\\
Certificate semantics & \path{ndnhw/certificates.py} & same-action pair counts, top-$k$ selection, exact/certificate/fallback classification\\
BPC-Greedy & \path{ndnhw/cert_bank_repair.py} & certificate-aware prefix selection across budget splits\\
Portfolio & \path{ndnhw/cert_bank_portfolio.py} & evaluates BPC, repair, exact-heavy, and shallow-heavy constructions\\
Real-name experiment & \path{experiments/run_v044_trace_replay_parallel.py} & deterministic sampling, parallel sweep, CSV/metadata export\\
\bottomrule
\end{tabular}
\label{tab:impl}
\end{table}

This suggests a stronger future optimization. Rather than using a constant $\gamma$, assign every potential certificate a physical cost $c_{s,t}$ that includes metadata, guard bits, and expected access cost:
\begin{equation}
|S|+\sum_{(s,t)\in K}c_{s,t}\le M.\tag{17}
\end{equation}
The algorithmic target becomes: \emph{find high-traffic action-equivalent classes with cheap one-sided signatures}. This directly connects semantic opportunity to implementable hardware cost.

\section{Implementation Correspondence and Reproducibility}
The implementation is a Python package plus experiment drivers. Table~\ref{tab:impl} maps the mathematical concepts to the released code structure used for the experiments.

The trace construction first inserts FIB prefixes, then inserts observed Interest paths, then replays every Interest against the full FIB trie to obtain its true LPM. The resulting true-LPM counts become FIB weights and, in the current bank model, access-load proxies. The repository test suite contains 13 automated tests covering trie/name behavior, certificate logic, bank repair, bank oracle cases, BPC, portfolio, PIT behavior, DP metrics, and structured traces.

The reported real-name run uses deterministic sample seed 1. Raw per-configuration rows, summary CSVs, metadata JSON, and source totals are preserved. In the final sweep, all 18 BPC-Portfolio winners report source BPC-Repair, i.e., the legacy code label for the BPC-Greedy branch.

\begin{table}[t]
\centering
\caption{Main real-name experiment configuration.}
\begin{tabular}{ll}
\toprule
Parameter & Value\\
\midrule
sample size / seed & 50,000 / 1\\
matched Interests & 49,961\\
parser-visible matched Interests at $D=8$ & 34,076\\
parser depth & $D=8$\\
object budgets & $M\in\{16,32,48\}$\\
bank counts & $B\in\{1,2,4\}$\\
bank regimes & fixed-total, fixed-per-bank\\
certificate cost & $\gamma=1.0$\\
headliner methods & BankRepairOnly, BPC-Greedy, BPC-Portfolio\\
\bottomrule
\end{tabular}
\label{tab:config}
\end{table}

\section{Evaluation Methodology}
\subsection{Workloads}
We use two complementary workloads. First, a synthetic structured trace isolates parser visibility: 1,200 Interests, 48 FIB prefixes, seed 7, and Zipf parameter $\alpha=1.1$. Second, we replay public NDN testbed names from the TNTech-NGIN dataset \cite{tntech}. The combined public data contains 265,925 names; the reported run samples 50,000 with seed 1, of which 49,961 have a matched FIB LPM.

The public packet traces do not contain a complete forwarding-state snapshot. We therefore derive a pruned FIB from observed prefixes at depths 1--4 and assign each retained prefix an action class from its first name component. This yields 143 FIB prefixes and six synthetic action classes; the sampled trie contains 14,976 nodes. At $D=8$, 34,076 matched Interests are parser-visible. The experiment is therefore a real-name replay under a reproducible synthetic forwarding-action snapshot, not a replay of a live router's routing state.

\subsection{Configuration}
The two bank regimes separate different abstract capacity assumptions. In fixed-total, increasing $B$ partitions fixed total capacity. In fixed-per-bank, each bank retains fixed capacity, so total capacity grows with $B$.

\subsection{Metrics}
Certified coverage counts exact hits plus accepted same-action certificate hits. All remaining traffic falls back. The would-be unsafe rate is the subset of fallback that matches selected shallow state but is neither exact nor certified. Prefix-bank feasibility is reported separately.

\begin{figure}[t]
\centering
\includegraphics[width=0.70\linewidth]{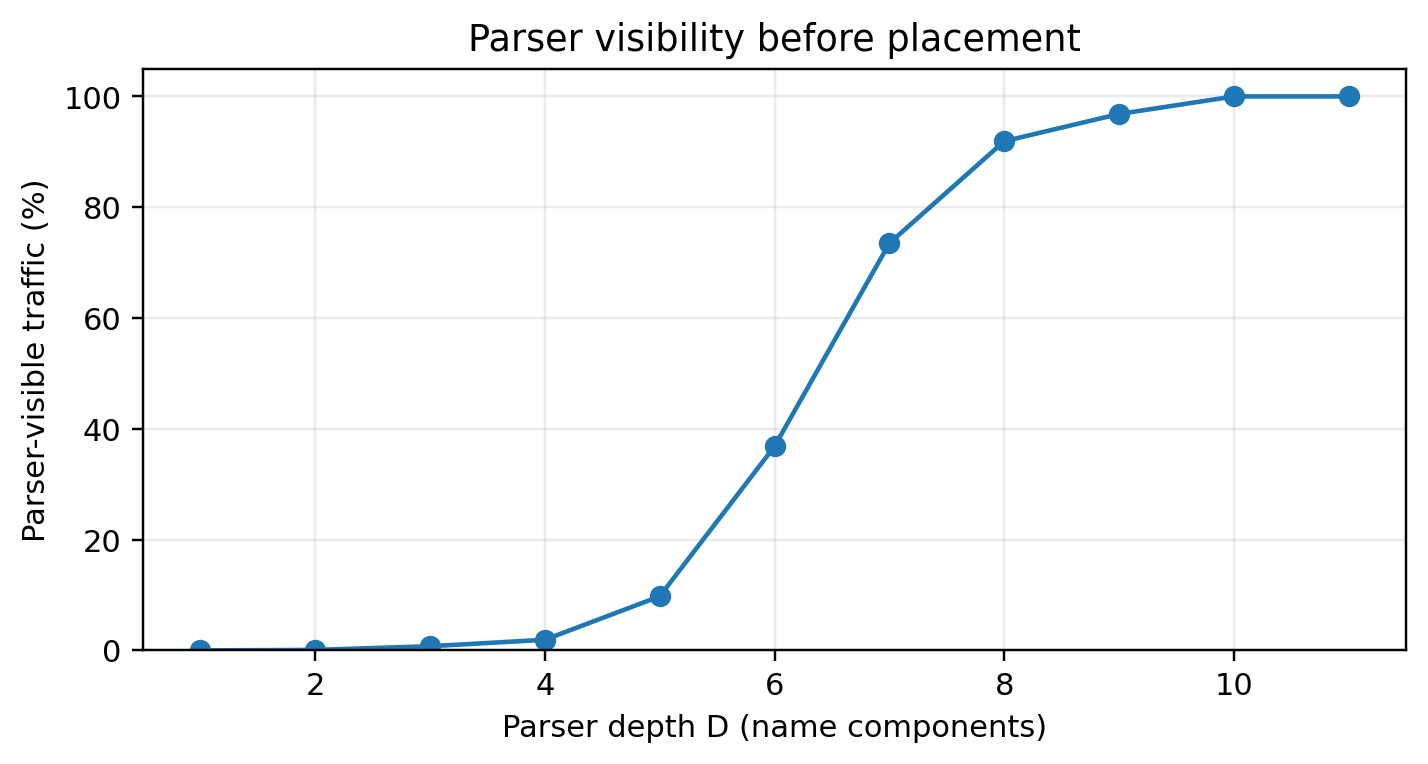}
\caption{Parser visibility on the 1,200-Interest synthetic sweep.}
\label{fig:visibility}
\end{figure}

\begin{table}[t]
\centering
\caption{Main modeled coverage, fallback (FB), and would-be unsafe (WU) rates; percentages.}
\begin{tabular}{c*{3}{r}*{3}{r}}
\toprule
& \multicolumn{3}{c}{BankRepairOnly} & \multicolumn{3}{c}{BPC-Portfolio}\\
\cmidrule(lr){2-4}\cmidrule(lr){5-7}
$M$ & Cov. & FB & WU & Cov. & FB & WU\\
\midrule
16 & 23.36 & 76.64 & 35.04 & 39.45 & 60.55 & 28.70\\
32 & 36.45 & 63.55 & 31.66 & 55.26 & 44.74 & 12.88\\
48 & 46.48 & 53.52 & 21.63 & 64.06 & 35.94 & 4.04\\
\bottomrule
\end{tabular}
\label{tab:coverage}
\end{table}

\section{Results}
\subsection{Parser depth is a first-order constraint}
Fig.~\ref{fig:visibility} shows the synthetic parser sweep. Parser-visible traffic rises from 36.83\% at $D=6$ to 91.92\% at $D=8$, reaching 100\% at $D=10$. This effect occurs before placement: a prefix or certificate cannot help a packet whose necessary name structure is not visible to the bounded parser.

\subsection{Certificate-aware placement increases modeled coverage}
At $M=16$, the best exact-only BankRepairOnly coverage is 23.36\%, while BPC-Portfolio reaches 39.45\%. At $M=32$, the values are 36.45\% and 55.26\%. At $M=48$, coverage rises from 46.48\% to 64.06\%, an absolute gain of 17.58 percentage points (approximately 37.8\% relative).

The selected BPC states at $M=16,32,48$ contain respectively 4, 8, and 16 trie nodes with certificate budgets 12, 24, and 32. At $M=48$, the normalized budget is therefore
\begin{equation}
16+1\cdot32=48.\tag{18}
\end{equation}
This split should not be read as a physical claim that one certificate occupies exactly the same number of bits as one trie node; $\gamma=1$ is the reference normalization used in this experiment.

\begin{figure}[t]
\centering
\includegraphics[width=0.70\linewidth]{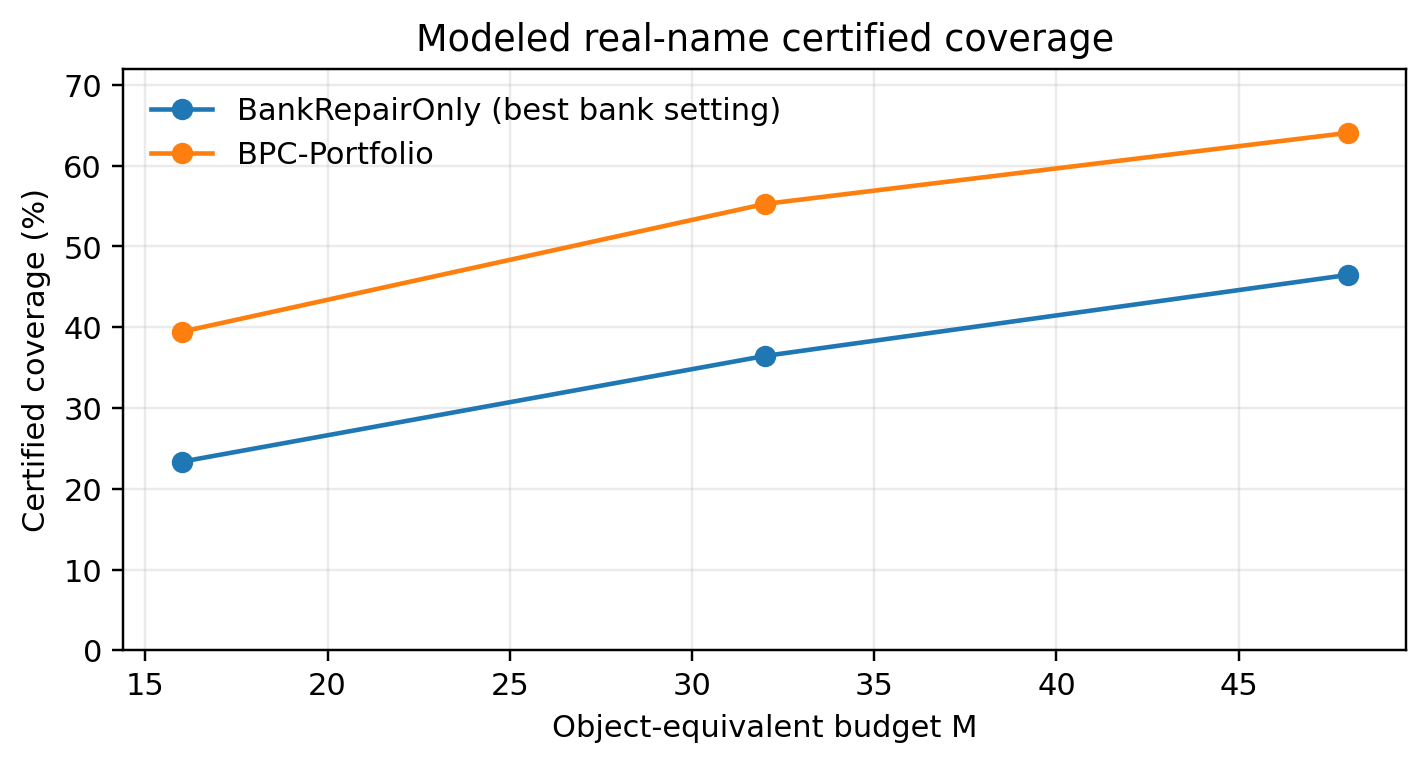}
\caption{Modeled certified coverage on the 50k real-name replay under the synthetic action snapshot. BankRepairOnly is the best exact-only result across tested bank settings at each budget.}
\label{fig:coverage}
\end{figure}

\subsection{The bank results require careful interpretation}\label{sec:banks}
\begin{figure}[t]
\centering
\includegraphics[width=0.70\linewidth]{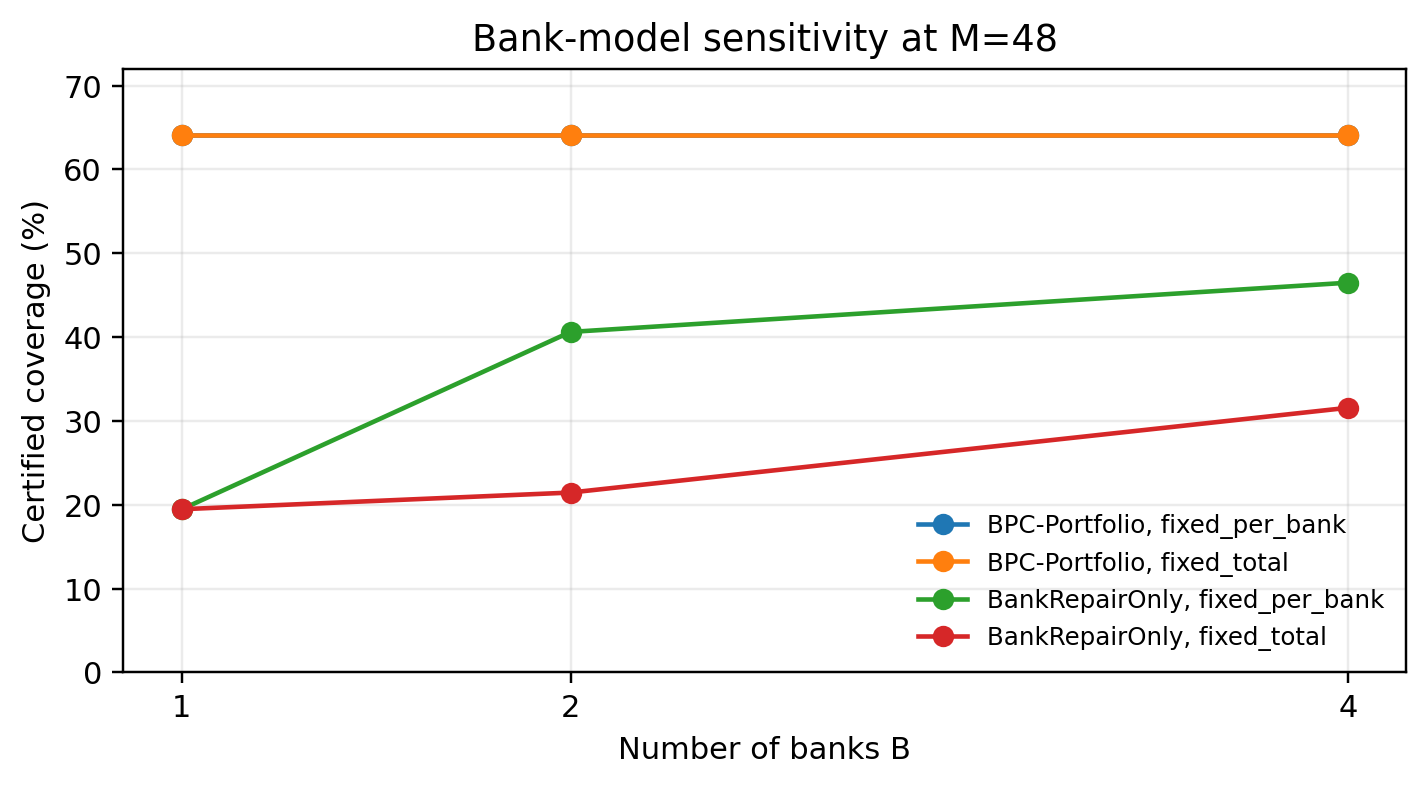}
\caption{Bank-model sensitivity at $M=48$. The flat BPC curve is a property of the current prefix-load proxy and selected state, not evidence of physical bank insensitivity.}
\label{fig:banks}
\end{figure}

Fig.~\ref{fig:banks} reproduces the bank-model sensitivity at $M=48$. Exact-only coverage is strongly affected by the tested regime, whereas BPC remains at 64.06\% because its selected prefix state is small enough to be feasible in every tested setting.

This result is intentionally interpreted conservatively. The current bank model has three important simplifications:
\begin{enumerate}
\item only selected trie nodes are assigned bank load; certificate metadata and guard accesses are not;
\item a shallow selected node is charged its trace-derived true-LPM load, not all descendant packets that may logically use it through certificates;
\item the bank assignment is an offline feasibility partition; runtime addressing, parallel probing, arbitration, and deepest-match reduction across banks are not simulated.
\end{enumerate}
Consequently, the bank experiments are best viewed as one abstract resource-pressure dimension of the placement study. A bank-free sensitivity run and a concrete certificate-table datapath are natural follow-up evaluations; their results cannot be inferred from the present CSVs without rerunning the mapper.

\section{When Certificates Are Most Valuable}
Certificates are most attractive in nearly action-homogeneous regions: a shallow prefix shares its action with several high-traffic deeper LPM classes, but the subtree contains enough conflicting exceptions that blindly collapsing the whole subtree would be incorrect.

Suppose $s$ has action A and descendants $t_1,t_2,t_3,t_4$ with actions A, A, A, B and traffic weights 30, 20, 15, 2. Storing $s$ plus three same-action certificates can potentially cover 65 units of descendant traffic while leaving the conflicting B class to fallback. By contrast, if every descendant has action A, a separate FIB-compression stage could potentially collapse the region more directly; if almost every descendant has a different action, certificate opportunity is small.

This motivates a useful future value model:
\begin{equation}
\operatorname{value}(s,t)\approx\frac{\text{traffic safely certified for }t}{\text{certificate bits}+\text{guard cost}+\text{expected access cost}}.\tag{19}
\end{equation}
The most interesting classes are not merely same-action classes; they are high-traffic same-action classes that admit cheap one-sided guards.

\section{Related Work}
High-speed NDN forwarding has been explored in software forwarders such as NFD and NDN-DPDK \cite{nfd,ndndpdk}, specialized name lookup structures \cite{wang2011,wang2012,so2013}, FPGA FIB lookup engines \cite{yu2019}, programmable-switch systems such as Vision and Pegasus \cite{vision,pegasus}, and probabilistic summaries such as adaptive prefix Bloom filters \cite{quan2014}. These systems demonstrate concrete fast paths on different execution substrates.

Our question is complementary: given a workload and explicit abstract resource budgets, which exact-LPM classes are safely representable by selected trie state plus semantics-preserving shortcuts? The focus is therefore offline placement and design-space characterization rather than an end-to-end forwarding datapath.

\section{Limitations and Research Directions}
The strongest limitations are architectural rather than algorithmic bookkeeping.

\textbf{Oracle-style guard.} The evaluator uses the known true LPM to instantiate the target-class guard. A practical design must replace this oracle with a bounded, sound implementation. The promising direction is not necessarily another complete LPM engine, but cheap one-sided signatures that certify profitable classes and reject uncertainty to fallback.

\textbf{Synthetic forwarding actions.} Public traces provide real names but not a complete live FIB snapshot. The first-component action assignment creates broad equivalence regions and can favor certificate opportunities. Action-model sensitivity and live forwarding-state snapshots are necessary before making claims about operational routing policies.

\textbf{Bank abstraction.} Prefix load is a true-LPM-derived proxy; certificates and guards carry no bank load; runtime bank lookup is not modeled. The current results therefore do not establish memory-system throughput.

\textbf{Normalized certificate cost.} The main result uses $\gamma=1$. A physical model should derive certificate and guard cost from bits, area, access energy, and bandwidth rather than treating all metadata objects as equal.

\textbf{No RTL/calibrated performance.} There is no synthesis, place-and-route, cycle timing, power, area, or calibrated throughput result. The contribution is an offline placement formulation and opportunity characterization.

\textbf{Algorithmic scaling.} Naive certificate-aware candidate evaluation repeatedly scans the trace. Incremental counts, lazy evaluation, subtree compression, and guard-aware pruning are promising engineering improvements.

\section{Conclusion}
This work reframes bounded NDN forwarding lookup as a combinatorial placement problem with a strict semantic boundary: a fast-path decision is accepted only when it preserves the exact forwarding action. Prefix closure produces reusable tree cost; exact memory-only placement admits dynamic programming; fixed-state certificate selection collapses to an exact top-$k$ problem; bank feasibility adds a load-partition layer; and the full difficulty comes from coupling these pieces because changing selected state changes the deepest hardware match and therefore the certificate universe.

BPC-Greedy exploits this coupling by scoring prefix closures using certificate-aware marginal safe coverage, while BPC-Portfolio hedges among certificate-aware, exact-heavy, repair-based, and shallow-heavy search biases. On the reported real-name replay under a reproducible synthetic action snapshot, the resulting placement increases modeled certified coverage from 46.48\% to 64.06\% at $M=48$.

The most important next step is now sharply defined: move from oracle certificates to implementable guarded certificates. In algorithmic terms, the next problem is to discover and co-optimize cheap, one-sided signatures for high-value action-equivalent classes, together with their physical storage, bank traffic, and latency. That extension would connect the placement opportunity exposed here to a complete hardware datapath.

\appendix
\section{Worked Microexample}
Consider FIB prefixes $s,t_1,t_2,t_3$ with $s\preceq t_i$, actions
\[
a(s)=A,\qquad a(t_1)=A,\qquad a(t_2)=A,\qquad a(t_3)=B,
\]
and exact-LPM trace weights 40, 25, and 20 for $t_1,t_2,t_3$. Assume selecting $s$ requires one new trie node and the certificate budget is two.

With $S=\{s\}$, the candidate certificate weights are
\[
q_S(s,t_1)=40,\qquad q_S(s,t_2)=25,
\]
while $(s,t_3)$ is invalid because actions differ. Top-2 certificate selection therefore certifies 65 packets. If another selected node $s'$ deeper in the $t_1$ branch is later added, packets formerly contributing to $(s,t_1)$ may instead become exact or contribute to a new pair $(s',t_1)$. This illustrates why certificate selection is trivial for fixed $S$ but prefix selection is not.

\section{Implementation-Level Pseudocode for Certificate Evaluation}
The evaluator logically performs the following classification for each trace Interest:
\begin{lstlisting}[basicstyle=\ttfamily\small,frame=single]
selected_match = deepest_selected_match(name, S)
true_lpm = lpm(name)

if selected_match == true_lpm:
    exact_hit
elif (selected_match, true_lpm) in chosen_certificates:
    certificate_hit
else:
    fallback
\end{lstlisting}
The second branch is where the current offline evaluator uses the trace-known true LPM as the guard oracle. A physical implementation must replace that membership test with a sound runtime guard.

\section{Why Certificates Are Not Prefix-Closed}
Prefix closure applies to $S$, because $S$ represents explicit trie state. The certificate set $K$ is metadata over semantic relations. A target $t$ can remain completely unselected while $(s,t)$ is stored, provided $s\in S\cap P$, $s\preceq t$, and $a(s)=a(t)$. Requiring certificate targets to be selected would defeat the main storage advantage of the construction.

\end{document}